\documentclass[twocolumn,final,natbib]{svjour3}         

\usepackage{graphicx}
\usepackage{mathptmx}

\newcommand{\Lsun}{\mbox{$L_\odot$}}
\newcommand{\kms}{\mbox{km\,s$^{-1}$}}
\newcommand{\micron}{\mbox{$\mu$m}}

\newcommand{\Lir}{\mbox{$L_{\rm{ir}}$}}          
\newcommand{\Lfir}{\mbox{$L_{\rm{fir}}$}}
\newcommand{\Lco}{\mbox{$L_{\rm{CO}}$}}
\newcommand{\Lhcn}{\mbox{$L_{\rm{HCN}}$}}

\newcommand{\HCN}{HCN(1--0)}
\newcommand{\HCO}{HCO$^{+}$(1--0)}

\newcommand{\aj}{AJ}                                  
\newcommand{\araa}{ARA\&A}                            
\newcommand{\apj}{ApJ}                                
\newcommand{\apjl}{ApJ}                               
\newcommand{\apjs}{ApJS}                              
\newcommand{\aap}{A\&A}                               
\newcommand{\mnras}{MNRAS}                            
\newcommand{\nat}{Nature}                             

\journalname{Astrophysics and Space Science}

\begin{document}

\title{Dense molecular gas in a sample of LIRGs and ULIRGs:}

\subtitle{The low-redshift connection to the huge high-redshift starbursts and AGNs}

\author{Javier Graci{\'a}-Carpio \and Santiago Garc{\'{\i}}a-Burillo \and Pere Planesas}

\institute{J. Graci{\'a}-Carpio \and S. Garc{\'{\i}}a-Burillo \and P. Planesas
           \at Observatorio Astron{\'o}mico Nacional \\ Alfonso XII, 3 \\ 28014 Madrid, Spain\\
           Tel.: +34-91-527-01-07 \\
           Fax:  +34-91-527-19-35\\
           \email{j.gracia@oan.es, s.gburillo@oan.es, p.planesas@oan.es}} 

\date{Received: date / Accepted: date}

\maketitle

\begin{abstract}
The sample of nearby LIRGs and ULIRGs for which dense molecular gas tracers have been measured is building up, allowing for the study of the physical and chemical properties of the gas in the variety of objects in which the most intense star formation and/or AGN activity in the local universe is taking place. This characterisation is essential to understand the processes involved, discard others and help to interpret the powerful starbursts and AGNs at high redshift that are currently being discovered and that will routinely be mapped by ALMA. We have studied the properties of the dense molecular gas in a sample of 17 nearby LIRGs and ULIRGs through millimeter observations of several molecules (HCO$^{+}$, HCN, CN, HNC and CS) that trace different physical and chemical conditions of the dense gas in these extreme objects. In this paper we present the results of our HCO$^{+}$ and HCN observations. We conclude that the very large range of measured line luminosity ratios for these two molecules severely questions the use of a unique molecular tracer to derive the dense gas mass in these galaxies. 
\end{abstract}

\section{Introduction \label{Introduction}}

The origin (starburst and/or AGN) of the infrared luminosity in luminous and ultraluminous infrared galaxies (LIRGs: $10^{11}\,\Lsun \leq \Lir < 10^{12}\,\Lsun$ and ULIRGs: $\Lir \geq 10^{12}\,\Lsun$) has been the subject of intense debate since their discovery \citep{Sanders88,Genzel98,Veilleux99,Gao04b}. In order to derive the dominant contribution to their IR luminosities the molecular gas properties of these galaxies have been extensively analysed through millimeter observations \citep[see the review of][]{Sanders96}. The higher star formation efficiency of the molecular gas (SFE $\propto \Lir/\Lco$) observed in LIRGs and ULIRGs compared to that found in spiral galaxies, led \citet{Sanders91} to propose that a dust enshrouded AGN contribute significantly to \Lir\ in these galaxies. However, \citet{Solomon92}, and more recently \citet{Gao04a}, found that the $\Lir/\Lhcn$ luminosity ratio, considered as a measure of the SFE of the dense gas traced by the \HCN\ transition, is almost constant independently of \Lir. According to this result the dense molecular gas properties in LIRGs and ULIRGs are similar to those in normal spiral galaxies and, as a result, the contribution to \Lir\ from a dust enshrouded AGN in LIRGs and ULIRGs is not required. These conclusions strongly depend on the assumption that \Lhcn\ is an unbiased tracer of dense molecular gas mass.  
 
\begin{figure*}[ht]
  \begin{center}
    \(\begin{array}{c@{\hspace{1cm}}c}
      \includegraphics[width=0.45\textwidth]{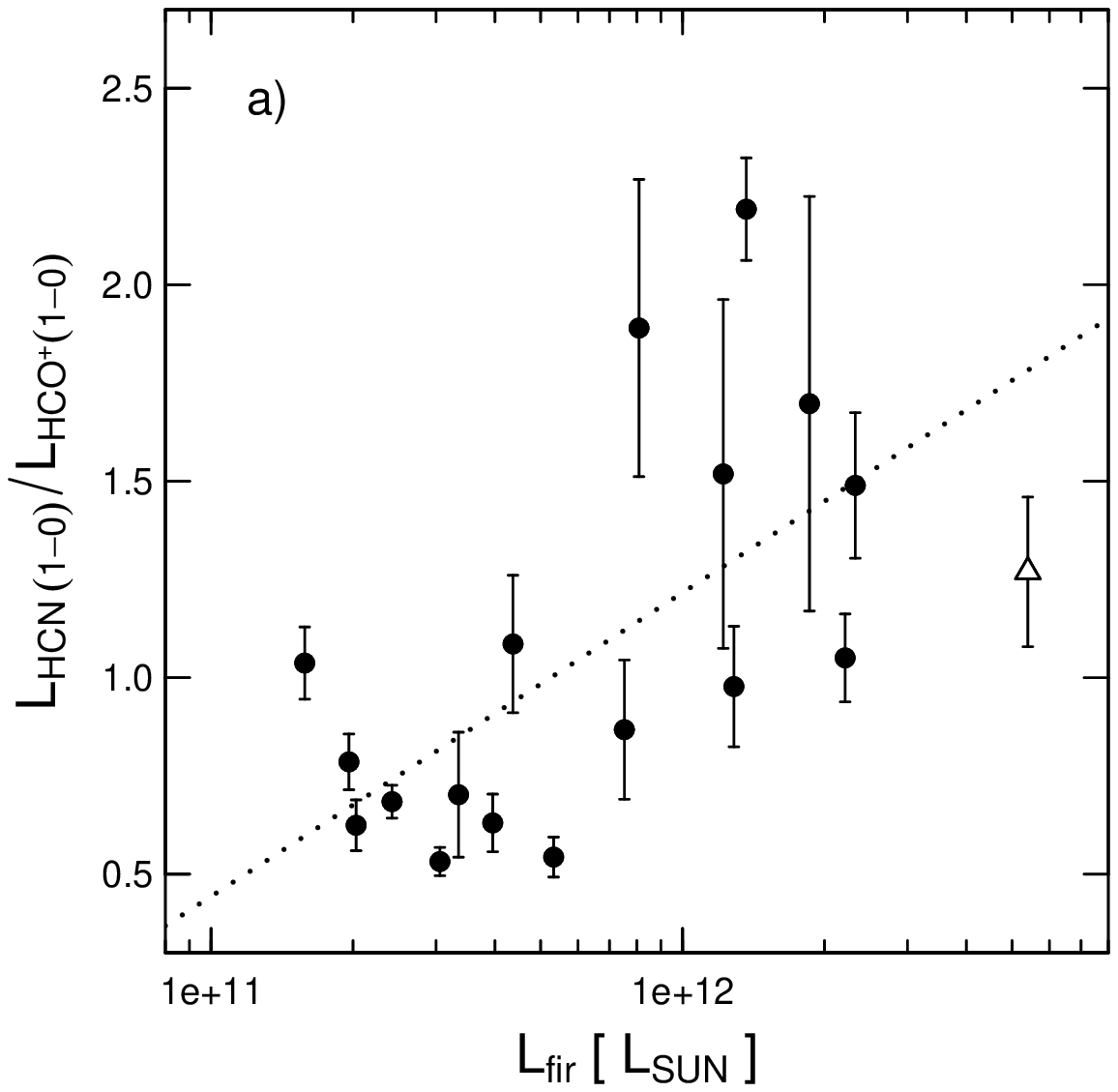} &
      \includegraphics[width=0.45\textwidth]{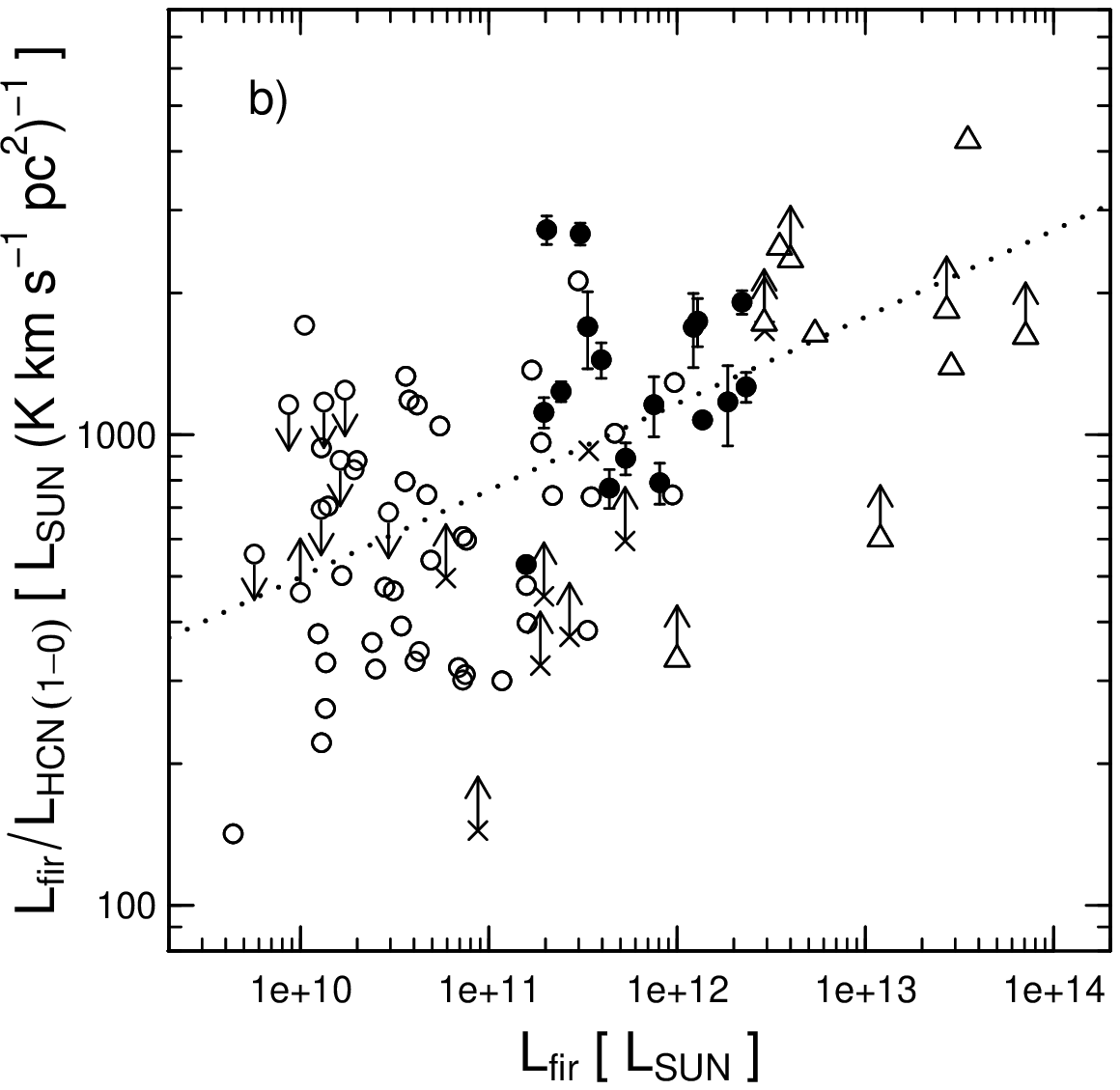} \\
    \end{array}\)
  \end{center}
\caption{{\bf a)} \HCN$/$\HCO\ luminosity ratio as a function of \Lfir\ in our sample of LIRGs and ULIRGs (black circles) and the Cloverleaf quasar \citep[open triangle;][]{Solomon03,Riechers06}. The observed trend indicates that the excitation and/or chemical properties of the dense molecular gas are different for LIRGs and ULIRGs (see \citealt{Gracia-Carpio06} for a more extended discussion). {\bf b)} $\Lfir/\Lhcn$ luminosity ratio as a function of \Lfir\ in a sample of normal spiral galaxies, LIRGs and ULIRGs \citep[open circles;][]{Gao04a}, our sample of LIRGs and ULIRGs (black circles), a sample of infrared-excess Palomar-Green QSOs \citep[crosses;][]{Evans06} and a sample of high-redshift objects \citep[open triangles;][and references therein]{Greve06}. Contrary to previous results \citep{Gao04b}, our new \HCN\ reobservations point to a significantly higher SFE of the dense gas for LIRGs and ULIRGs compared to normal less luminous spiral galaxies. Palomar-Green QSOs and high-redshift objects fall within the linear regression fit (dotted line) calculated for our sample of LIRGs and ULIRGs and \citet{Gao04a} galaxies. The higher SFE traced by \Lhcn\ in LIRGs and ULIRGs may be related to a real difference of the dense molecular gas properties in these galaxies or to an additional contribution to \Lfir\ from a dust enshrouded AGN.}
\label{figure 1}
\end{figure*}
 
However, recent results have cast several doubts about the reliability of HCN as an unbiased tracer of the dense molecular gas mass in galaxies \citep{Kohno01,Usero04,Kohno05}, being LIRGs, ULIRGs and high-redshift galaxies a particular case \citep{Gracia-Carpio06,Imanishi06,Garcia-Burillo06}. The main concerns about the use of \Lhcn\ as a quantitative probe of the dense gas mass come from the particular chemistry and excitation conditions of this molecule. HCN abundance can be significantly enhanced under the influence of X-ray chemistry driven by an embedded AGN \citep{Lepp96,Maloney96} or in the molecular gas closely associated with high-mass star forming regions \citep{Blake87,Lahuis06}. In addition to that, the excitation of HCN lines might be affected by IR pumping through a 14\,\micron\ vibrational transition near strong mid-infrared sources \citep{Aalto95}. All these effects will contribute to increase the total \HCN\ emission, breaking the claimed proportionality between \Lhcn\ and the total dense molecular gas mass. In this context, the conclusions extracted by \citet{Gao04b} about the starburst origin of the IR luminosity in LIRGs and ULIRGs can be questioned.

\section{Sample selection and observations \label{Sample selection}}

In order to test if the \HCN\ emission is a fair tracer of the dense molecular gas mass we have conducted with the IRAM 30-meter telescope a dense molecular gas survey in a sample of 17 LIRGs and ULIRGs selected to cover homogeneously the \Lir\footnote{All luminosities have been calculated assuming a flat $\rm{\Lambda}$-dominated cosmology described by $\rm{H_{0}} = 71\,\kms\,\rm{Mpc}^{-1}$ and $\rm{\Omega_{m}} = 0.27$ \citep{Spergel03}.} range between $10^{11.3}\,\Lsun$ and $10^{12.5}\,\Lsun$. All galaxies are located at distances larger than 50\,Mpc to be confident that the total emission of the molecular gas can be measured in a single pointing (\HCO\ beam $\sim 28'' = 7\,\rm{kpc}$ at 50\,Mpc). Several molecules (HCO$^{+}$, HCN, CN, HNC and CS) and rotational transitions (J=1--0, 3--2) were observed in 7 periods between November 2004 and November 2006. The results from our full observations will be discussed in a future paper. Here we will concentrate on our HCO$^{+}$ and HCN results.

\section{Dense molecular gas in LIRGs and ULIRGs \label{LIRGs and ULIRGs}}

\begin{figure*}
  \begin{center}
    \(\begin{array}{c@{\hspace{1cm}}c}
      \includegraphics[width=0.45\textwidth]{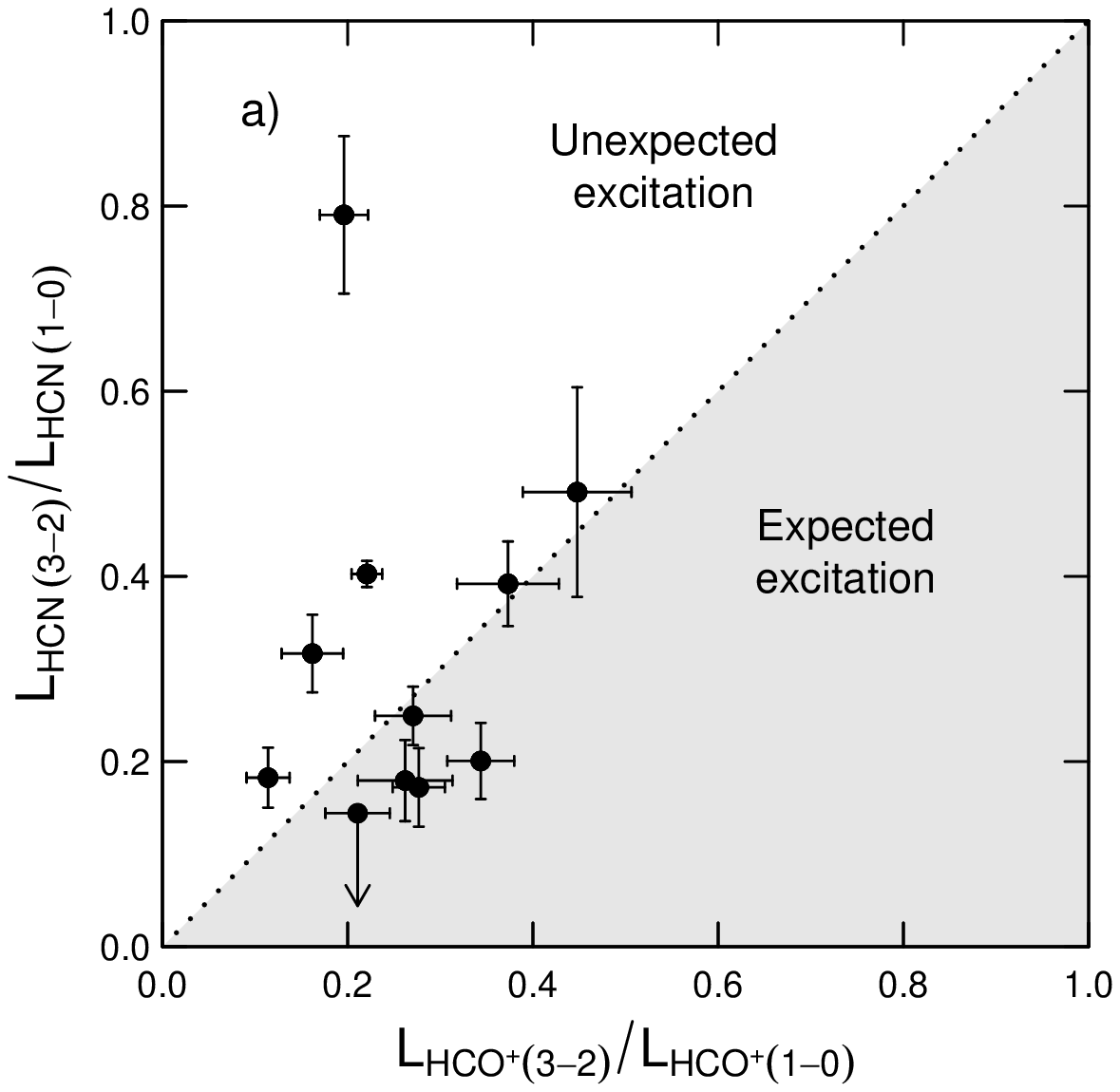} &
      \includegraphics[width=0.45\textwidth]{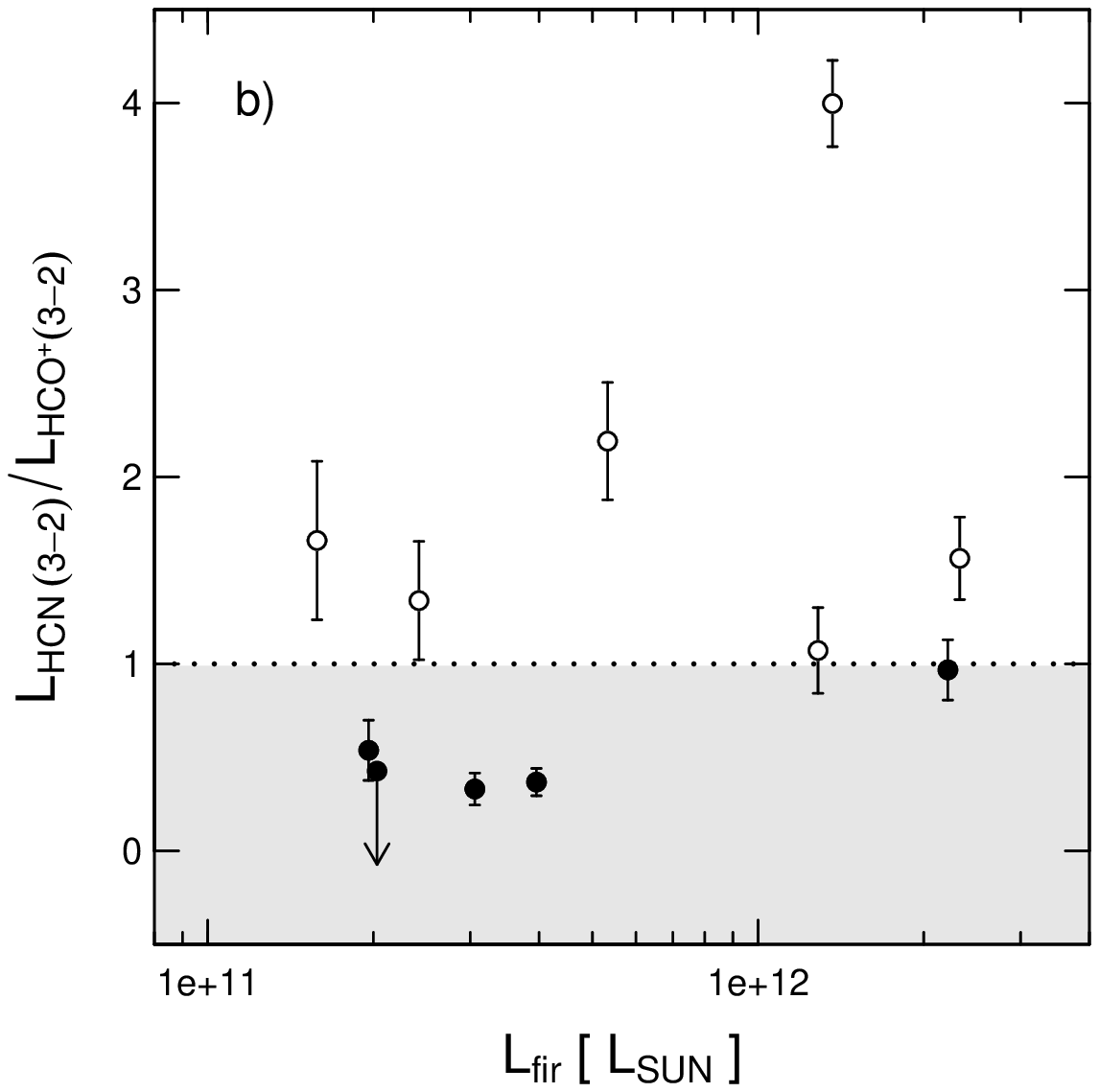} \\
    \end{array}\)
  \end{center}
\caption{{\bf a)} HCN(3--2)$/$HCN(1--0) luminosity ratio against the HCO$^{+}$(3--2)$/$HCO$^{+}$(1--0) luminosity ratio in our sample of LIRGs and ULIRGs. We can see that HCN and HCO$^{+}$ excitation is clearly subthermal for most galaxies. We have highlighted in grey the region of `expected excitation' predicted by our simple LVG calculations assuming similar abundances for both molecules (see text). About half of the galaxies show an `unexpected excitation' that can be explained if $\rm [HCN]/[HCO^{+}] > 10$. {\bf b)} HCN(3--2)$/$HCO$^{+}$(3--2) luminosity ratio as a function of \Lfir\ in our sample of LIRGs and ULIRGs. We have highlighted in grey the region where HCN(3--2)$/$HCO$^{+}$(3--2) $\leq$ 1, as predicted by our LVG calculations. To explain the J=3--2 line emission in those galaxies that do not satisfy the mentioned condition we need HCN to be overabundant with respect to HCO$^{+}$. Open circles represent those galaxies with `unexpected excitation' in {\bf a)}.}
\label{figure 2}
\end{figure*}

In a recent article \citep{Gracia-Carpio06} we presented the results of our HCO$^{+}$ survey in LIRGs and ULIRGs and showed an intriguing trend between the HCN$/$HCO$^{+}$ J=1--0 luminosity ratio and \Lir. In that paper we discussed that the observed trend could be the result of an anomalous excitation and/or chemistry of the HCN molecule \citep[or, alternatively, of the HCO$^{+}$ molecule; see also][]{Papadopoulos06}. In Fig.\@~\ref{figure 1}a) we show an updated version of the same plot including our recent \HCN\ reobservations of the \citet{Solomon92} sample. With the addition of the new data the existence of a trend is confirmed. Independently of the origin of this trend (see \citealt{Gracia-Carpio06} for a detailed discussion), it is evident that the properties of the dense molecular gas in LIRGs are different from those in ULIRGs. It is also clear that it is necessary to observe several dense gas tracers in order to characterise the molecular gas properties in these galaxies.

Fig.\@~\ref{figure 1}b) also shows that dense molecular gas properties seem to change with increasing \Lfir\ from normal spiral galaxies to LIRGs, ULIRGs and high-redshift galaxies. We can see that the SFE of the dense gas traced by the \HCN\ line luminosity is significantly higher in LIRGs and ULIRGs than in normal spiral galaxies. We can interpret this result in three different ways. First, it may represent a real variation of the SFE of the molecular gas with \Lfir; this is not surprising as the Initial Mass Function (IMF) and the Schmidt Law do not need to be the same in different extragalactic environments. Second, if we assume that the SFE of the dense gas is constant independently of \Lfir, it may indicate that the SFE should be measured using a different tracer of the dense gas mass with a higher critical density than \HCN. Third, Fig.\@~\ref{figure 1}b) may indicate that there is an additional contribution to \Lfir\ from a dust enshrouded AGN in the most IR luminous galaxies. This is probably true in the case of Palomar-Green QSOs and high-redshift galaxies \citep{Rowan-Robinson00}, but it is not clear in the case of LIRGs and ULIRGs. We should note that if \Lhcn\ overestimates the dense molecular gas content at high IR luminosities, this would imply that the reported increase of the SFE of the dense gas as a function of \Lfir\ would be even higher. 

In an effort to constrain the relative abundances and excitation properties of HCN and HCO$^{+}$ we have studied their J = 3--2$/$1--0 luminosity ratios in Fig.\@~\ref{figure 2}a). The critical densities of HCN rotational transitions are higher than those of HCO$^{+}$ by a factor of $\sim 6$. That means that HCN and HCO$^{+}$ may trace different gas phases with different densities, and that if we want to model the rotational emission of these molecules we need to consider two molecular gas phases with different density and temperature. However, given the small number of line ratios available, we have taken a more simple approach to interpret our results and we have assumed that both molecules trace the same molecular gas phase (i.e. similar density and kinetic temperature). If we also assume an abundance ratio $\rm [HCN]/[HCO^{+}] \sim 1$, and that collisional excitation dominates the rotational emission of these molecules, then a simple LVG analysis indicates that the HCO$^{+}$ molecule should be more excited (i.e. a higher J = 3--2$/$1--0 ratio) than HCN. In Fig.\@~\ref{figure 2}a) we have highlighted in grey the region where HCO$^{+}$(3--2)$/$HCO$^{+}$(1--0) $>$ HCN(3--2)$/$HCN(1--0). About half of the galaxies fall into this region of `expected excitation'. However, the other half fall into a region that was not expected by our simple LVG calculations. This `unexpected excitation' can be explained if $\rm [HCN]/[HCO^{+}] \geq 10$ in these galaxies, an abundance ratio much higher than those observed in Galactic star-forming regions \citep[see for example Tab. 10 in][]{Stauber06}. This result does not depend on the density, kinetic temperature and column density of H$_{2}$ considered in the calculations. We note that in a two phase model the $\rm [HCN]/[HCO^{+}]$ abundance problem would still appear in the densest phase.

We have adopted a similar approach in Fig.\@~\ref{figure 2}b) where we have represented the HCN(3--2)$/$HCO$^{+}$(3--2) luminosity ratio as a function of \Lfir\ in our sample of LIRGs and ULIRGs. If we assume that both molecules trace the same molecular gas phase and have similar abundances, then the HCN(3--2)$/$HCO$^{+}$(3--2) luminosity ratio should be $< 1$. We have highlighted in grey the region where this condition is fulfilled. Again, half of the galaxies do not fall into the region predicted by our simple model. HCN(3--2)$/$HCO$^{+}$(3--2) $\geq$ 1 is predicted for $\rm [HCN]/[HCO^{+}] \geq 7$, independently of the density, kinetic temperature and H$_{2}$ column density of the medium.

\section{Dense molecular gas in high-redshift galaxies \label{high redshift}}

Observations of dense molecular gas at high redshift are still rare and difficult. Only four high-redshift galaxies have been detected in HCN to date \citep{Solomon03,Vanden-Bout04,Carilli05,Wagg05}, two in HCO$^{+}$ \citep{Riechers06,Garcia-Burillo06}
and one in HNC and tentatively in CN \citep{Guelin07}. In spite of the difficulties, it is possible to apply the same kind of analysis described above to high redshift objects. \citet{Wagg05}, \citet{Garcia-Burillo06} and \citet{Guelin07} have used simple radiative transfer calculations to impose stringent constraints on [HCN]$/$[CO], [HCN]$/$[HCO$^{+}$], [HCN]$/$[HNC] and [HCN]$/$[CN] abundance ratios in the broad absorption line quasar APM 08279+5255 at $z=3.9$. Their results point to HCN being overabundant with respect to HCO$^{+}$ and CN by a factor of $\sim 10$, while [HCN]$/$[CO] $\sim 10^{-2}$--$10^{-3}$ and [HCN]$/$[HNC] $\sim 1.6$. Infrared pumping through higher vibrational transitions may also play a role in the excitation of some of these molecular lines and can equally explain the observed luminosity ratios. The infrared luminosity of APM 08279+5255 is dominated by the contribution of its AGN \citep{Rowan-Robinson00}, which makes this galaxy an ideal candidate to test the effects of the feedback of activity in the properties of the dense molecular gas. The fact that in this galaxy HCN seems to be overabundant with respect to HCO$^{+}$, as we found in some of the galaxies of our sample of LIRGs and ULIRGs, could indicate that similar processes dominate the dense molecular gas chemistry in these galaxies.


\begin{thebibliography}{}

\bibitem[\protect\citeauthoryear{Aalto \etal}{1995}]{Aalto95}
Aalto, S., Booth, R.~S., Black, J.~H., Johansson, L.~E.~B.: 
\aap\ 300, 369 (1995)

\bibitem[\protect\citeauthoryear{Blake \etal}{1987}]{Blake87}
Blake, G.~A., Sutton, E.~C., Masson, C.~R., Phillips, T.~G.: 
\apj\ 315, 621 (1987)

\bibitem[\protect\citeauthoryear{Carilli \etal}{2005}]{Carilli05}
Carilli, C.~L., Solomon, P., Vanden Bout, P., Walter, F., Beelen, A., Cox, P., Bertoldi, F., Menten, K.~M., Isaak, K.~G., Chandler, C.~J., Omont, A.: 
\apj\ 618, 586 (2005)

\bibitem[\protect\citeauthoryear{Evans \etal}{2006}]{Evans06}
Evans, A.~S., Solomon, P.~M., Tacconi, L.~J., Vavilkin, T., Downes, D.: 
\aj\ 132, 2398 (2006)

\bibitem[\protect\citeauthoryear{Gao \&\ Solomon}{2004a}]{Gao04a}
Gao, Y., Solomon, P.~M.: 
\apjs\ 152, 63 (2004a)

\bibitem[\protect\citeauthoryear{Gao \&\ Solomon}{2004b}]{Gao04b}
Gao, Y., Solomon, P.~M.: 
\apj\ 606, 271 (2004b)

\bibitem[\protect\citeauthoryear{Garc{\'{\i}}a-Burillo \etal}{2006}]{Garcia-Burillo06}
Garc{\'{\i}}a-Burillo, S., Graci{\'a}-Carpio, J., Gu{\'e}lin, M., Neri, R., Cox, P., Planesas, P., Solomon, P.~M., Tacconi, L.~J., Vanden Bout, P.~A.: 
\apjl\ 645, L17 (2006)

\bibitem[\protect\citeauthoryear{Genzel \etal}{1998}]{Genzel98}
Genzel, R., Lutz, D., Sturm, E., Egami, E., Kunze, D., Moorwood, A.~F.~M., Rigopoulou, D., Spoon, H.~W.~W., Sternberg, A., Tacconi-Garman, L.~E., Tacconi, L., Thatte, N.: 
\apj\ 498, 579 (1998)

\bibitem[\protect\citeauthoryear{Graci{\'a}-Carpio \etal}{2006}]{Gracia-Carpio06}
Graci{\'a}-Carpio, J., Garc{\'{\i}}a-Burillo, S., Planesas, P., Colina, L.: 
\apjl\ 640, L135 (2006)

\bibitem[\protect\citeauthoryear{Greve \etal}{2006}]{Greve06}
Greve, T.~R., Hainline, L.~J., Blain, A.~W., Smail, I., Ivison, R.~J., Papadopoulos, P.~P.: 
\aj\ 132, 1938 (2006)

\bibitem[\protect\citeauthoryear{Gu{\'e}lin \etal}{2007}]{Guelin07}
Gu{\'e}lin, M., Salom{\'e}, P., Neri, R., Garc{\'{\i}}a-Burillo, S., Graci{\'a}-Carpio, J., Cernicharo, J., Cox, P., Planesas, P., Solomon, P.~M., Tacconi, L.~J., vanden Bout, P.: 
\aap\ 462, L45 (2007)

\bibitem[\protect\citeauthoryear{Imanishi \etal}{2006}]{Imanishi06}
Imanishi, M., Nakanishi, K., Kohno, K.: 
\aj\ 131, 2888 (2006)

\bibitem[\protect\citeauthoryear{Kohno \etal}{2001}]{Kohno01}
Kohno, K., Matsushita, S., Vila-Vilar{\'o}, B., Okumura, S.~K., Shibatsuka, T., Okiura, M., Ishizuki, S., Kawabe, R.: 
ASP Conf. Ser. 249: The Central Kiloparsec of Starbursts and AGN: The La Palma Connection,
672. Knapen, J.~H., Beckman, J.~E., Shlosman, I., \&\ Mahoney, T.~J., (2001)

\bibitem[\protect\citeauthoryear{Kohno}{2005}]{Kohno05}
Kohno, K., 
AIP Conf. Proc. 783: The Evolution of Starbursts,
203. H{\"u}ttmeister, S., Manthey, E., Bomans, D., \&\ Weis, K., (2005)

\bibitem[\protect\citeauthoryear{Lahuis \etal}{2006}]{Lahuis06}
Lahuis, F., Spoon, H.~W.~W., Tielens, A.~G.~G.~M., Doty, S.~D., Armus, L., Charmandaris, V., Houck, J.~R., St{\"a}uber, P., van Dishoeck, E.~F.:
astro-ph/0612748 (2006)

\bibitem[\protect\citeauthoryear{Lepp \&\ Dalgarno}{1996}]{Lepp96}
Lepp, S., Dalgarno, A.: 
\aap\ 306, L21 (1996)

\bibitem[\protect\citeauthoryear{Maloney \etal}{1996}]{Maloney96}
Maloney, P.~R., Hollenbach, D.~J., Tielens, A.~G.~G.~M.: 
\apj\ 466, 561 (1996)

\bibitem[\protect\citeauthoryear{Papadopoulos}{2006}]{Papadopoulos06}
Papadopoulos, P.~P.: 
astro-ph/0610477 (2006)

\bibitem[\protect\citeauthoryear{Riechers \etal}{2006}]{Riechers06}
Riechers, D.~A., Walter, F., Carilli, C.~L., Weiss, A., Bertoldi, F., Menten, K.~M., Knudsen, K.~K., Cox, P.: 
\apjl\ 645, L13 (2006)

\bibitem[\protect\citeauthoryear{Rowan-Robinson}{2000}]{Rowan-Robinson00}
Rowan-Robinson, M.: 
\mnras\ 316, 885 (2000)

\bibitem[\protect\citeauthoryear{Sanders \etal}{1988}]{Sanders88}
Sanders, D.~B., Soifer, B.~T., Elias, J.~H., Madore, B.~F., Matthews, K., Neugebauer, G., Scoville, N.~Z.: 
\apj\ 325, 74 (1988)

\bibitem[\protect\citeauthoryear{Sanders \etal}{1991}]{Sanders91}
Sanders, D.~B., Scoville, N.~Z., Soifer, B.~T.: 
\apj\ 370, 158 (1991)

\bibitem[\protect\citeauthoryear{Sanders \&\ Mirabel}{1996}]{Sanders96}
Sanders, D.~B., Mirabel, I.~F.: 
\araa\ 34, 749 (1996)

\bibitem[\protect\citeauthoryear{Solomon \etal}{1992}]{Solomon92}
Solomon, P.~M., Downes, D., Radford, S.~J.~E.: 
\apjl\ 387, L55 (1992)

\bibitem[\protect\citeauthoryear{Solomon \etal}{2003}]{Solomon03}
Solomon, P., Vanden Bout, P., Carilli, C., Guelin, M.: 
\nat\ 426, 636 (2003)

\bibitem[\protect\citeauthoryear{Spergel \etal}{2003}]{Spergel03}
Spergel, D.~N., Verde, L., Peiris, H.~V., Komatsu, E., Nolta, M.~R., Bennett, C.~L., Halpern, M., Hinshaw, G., Jarosik, N., Kogut, A., Limon, M., Meyer, S.~S., Page, L., Tucker, G.~S., Weiland, J.~L., Wollack, E., Wright, E.~L.: 
\apjs\ 148, 175 (2003)

\bibitem[\protect\citeauthoryear{St{\"a}uber \etal}{2006}]{Stauber06}
St{\"a}uber, P., Benz, A.~O., J{\o}rgensen, J.~K., van Dishoeck, E.~F., Doty, S.~D., van der Tak, F.~F.~S.: 
astro-ph/0608393 (2006)

\bibitem[\protect\citeauthoryear{Usero \etal}{2004}]{Usero04}
Usero, A., Garc{\'{\i}}a-Burillo, S., Fuente, A., Mart{\'{\i}}n-Pintado, J., Rodr{\'{\i}}guez-Fern{\'a}ndez, N.~J.: 
\aap\ 419, 897 (2004)

\bibitem[\protect\citeauthoryear{Vanden Bout \etal}{2004}]{Vanden-Bout04}
Vanden Bout, P.~A., Solomon, P.~M., Maddalena, R.~J.: 
\apjl\ 614, L97 (2004)

\bibitem[\protect\citeauthoryear{Veilleux \etal}{1999}]{Veilleux99}
Veilleux, S., Kim, D.-C., Sanders, D.~B.: 
\apj\ 522, 113 (1999)

\bibitem[\protect\citeauthoryear{Wagg \etal}{2005}]{Wagg05}
Wagg, J., Wilner, D.~J., Neri, R., Downes, D., Wiklind, T.: 
\apjl\ 634, L13 (2005)

\end{thebibliography}
\end{document}